\documentclass[sigplan,screen]{acmart}

\AtBeginDocument{%
  }

\copyrightyear{2023}
\acmYear{2023}
\setcopyright{rightsretained}
\acmConference[ISMM '23]{Proceedings of the 2023 ACM SIGPLAN International Symposium on Memory Management}{June 18, 2023}{Orlando, FL, USA}
\acmBooktitle{Proceedings of the 2023 ACM SIGPLAN International Symposium on Memory Management (ISMM '23), June 18, 2023, Orlando, FL, USA}
\acmDOI{10.1145/3591195.3595279}
\acmISBN{979-8-4007-0179-5/23/06}
\acmPrice{}
\received{2023-03-03}
\received[accepted]{2023-04-24}

\usepackage{amsmath}
\usepackage[normalem]{ulem}
\usepackage{hyperref}
\usepackage{listings}
\usepackage{graphicx}
\usepackage{tablefootnote}
\usepackage{caption}
\usepackage{subcaption}
\usepackage{multirow}
\usepackage{enumitem}
\usepackage{microtype}
\usepackage{balance}
\setlist{parsep=0pt,listparindent=\parindent}

\begin{document}

\title{The Unexpected Efficiency of Bin Packing Algorithms for Dynamic Storage Allocation in the Wild}
\subtitle{An Intellectual Abstract}

\author{Christos Lamprakos}
\email{cplamprakos@microlab.ntua.gr}
\orcid{0000-0002-3370-857X}
\affiliation{%
  \institution{National Technical University of Athens}
  \streetaddress{9, Iroon Polytechniou St.}
  \city{Athens}
  \country{Greece}
  \postcode{157 80}
}
\affiliation{%
  \institution{Katholieke Universiteit Leuven}
  \streetaddress{Oude Markt 13}
  \city{Leuven}
  \country{Belgium}
  \postcode{3000}
}

\author{Sotirios Xydis}
\orcid{0000-0003-3151-2730}
\email{sxydis@microlab.ntua.gr}
\affiliation{%
  \institution{National Technical University of Athens}
  \streetaddress{9, Iroon Polytechniou St.}
  \city{Athens}
  \country{Greece}
  \postcode{157 80}
}

\author{Francky Catthoor}
\orcid{0000-0002-3599-8515}
\email{francky.catthoor@imec.be}
\affiliation{%
  \institution{IMEC Science Park}
  \streetaddress{Gaston Geenslaan 14}
  \city{Leuven}
  \country{Belgium}
  \postcode{3001}
}
\affiliation{%
  \institution{Katholieke Universiteit Leuven}
  \streetaddress{Oude Markt 13}
  \city{Leuven}
  \country{Belgium}
  \postcode{3000}
}

\author{Dimitrios Soudris}
\orcid{0000-0002-6930-6847}
\email{dsoudris@microlab.ntua.gr}

\affiliation{%
  \institution{National Technical University of Athens}
  \streetaddress{9, Iroon Polytechniou St.}
  \city{Athens}
  \country{Greece}
  \postcode{157 80}
}

\renewcommand{\shortauthors}{Lamprakos, Xydis, Catthoor and Soudris}

\begin{abstract}
Two-dimensional rectangular bin packing (2DBP) is a known abstraction of dynamic storage allocation (DSA). We argue that such abstractions can aid practical purposes. 2DBP algorithms optimize their placements' makespan, i.e., the size of the used address range. At first glance modern virtual memory systems with demand paging render makespan irrelevant as an optimization criterion: allocators commonly employ sparse addressing and need worry only about fragmentation caused within page boundaries. But in the embedded domain, where portions of memory are statically pre-allocated, makespan remains a reasonable metric.

Recent work has shown that viewing allocators as black-box 2DBP solvers bears meaning. For instance, there exists a 2DBP-based fragmentation metric which often correlates monotonically with maximum resident set size (RSS). Given the field's indeterminacy with respect to fragmentation definitions, as well as the immense value of physical memory savings, we are motivated to set allocator-generated placements against their 2DBP-devised, makespan-optimizing counterparts. Of course, allocators must operate online while 2DBP algorithms work on complete request traces; but since both sides optimize criteria related to minimizing memory wastage, the idea of studying their relationship preserves its intellectual--and practical--interest.

Unfortunately no implementations of 2DBP algorithms for DSA are available. This paper presents a first, though partial, implementation of the state-of-the-art. We validate its functionality by comparing its outputs' makespan to the theoretical upper bound provided by the original authors. Along the way, we identify and document key details to assist analogous future efforts.

Our experiments comprise 4 modern allocators and 8 real application workloads. We make several notable observations on our empirical evidence: in terms of makespan, allocators outperform Robson's worst-case lower bound $93.75\%$ of the time. In $87.5\%$ of cases, GNU's \texttt{malloc} implementation demonstrates equivalent or superior performance to the 2DBP state-of-the-art, despite the second operating offline. 

Most surprisingly, the 2DBP algorithm proves competent in terms of fragmentation, producing up to $2.46$x better solutions. Future research can leverage such insights towards memory-targeting optimizations.
\end{abstract}




\maketitle

\section{Introduction}
\label{sec:intro}
Despite dynamic memory allocation's (DSA) omnipresence in modern computing, it is, in the context of first principles, mistreated. The fundamental enemy that is fragmentation has yet to receive a clear, quantitative definition~\cite{jemalloc, realfrag}. And despite knowing for decades the interplay between workload behavior, allocator policy and fragmentation~\cite{wilson1995surv}, we lack a systematic approach to characterize programs based on their dynamic memory characteristics. These gaps impose invisible costs to systems in terms of physical memory. 

In a sister work~\cite{lamprakos2023viewing}, we claim that constructing representations of workload-allocator interaction\footnote{In this text, by ``allocator'' we always mean general-purpose, non-moving allocators managing Linux virtual memory.} as instances of two-dimensional rectangular bin packing (2DBP) can serve as an informed basis for future analysis and design methods. We support our claim by defining a fragmentation metric in our representation space, and showing that, for many workloads, our metric correlates with maximum resident set size (RSS) in a monotonically increasing fashion. This intellectual abstract commences from where that work ends: \textit{if 2DBP is a potent substrate, how do real allocator placements compare to solutions produced by 2DBP algorithms?}

Such a comparison may seem counterintuitive at first; allocators operate online, while 2DBP algorithms take complete request sequences as input. But it is precisely this difference that makes our investigation worthwhile, for it offers an empirical view of allocators' practical limits. By setting allocators against offline oracles we can measure their distance from optimal behavior. This distance is expected to vary across different applications, and thus enables us to detect workloads that have a lot to gain from custom placement policies. In the opposite direction, one's search for custom policies may be inspired by work in the 2DBP field.

Several 2DBP subcategories exist, but only a specific one is suitable for DSA~\cite{chrobak1988some}. Rectangles are often allowed to slide in both dimensions of the 2D plane~\cite{jylanki2010thousand}; in our case, rectangle position on one axis must be fixed, denoting allocation and deallocation time respectively. This variation is commonly referred to as DSA in the theoretical literature~\cite{buchsbaum}. To the best of our knowledge, there are no implementations of 2DBP DSA algorithms available. We thus embarked on implementing the state-of-the-art, published by Buchsbaum et al. in 2003~\cite{buchsbaum}. This paper records all insights gained along the way. We make the following contributions:

\begin{itemize}
    \item a partial implementation of the state-of-the-art 2DBP algorithm\footnote{From this point onwards, we are going to abbreviate said algorithm as BA, i.e., ``the algorithm published in 2003 by Buchsbaum et al.''~\cite{buchsbaum}} suitable for modeling DSA
    \item a set of remarks on shortcomings of both BA itself and our own implementation\footnote{In reality we cannot be certain about the observed shortcomings' cause. It could as much be the case that our source code contains unidentified bugs, as that transitioning from purely theoretical constructs to practical implementations is expected to yield difficulty.
    
    We are indebted to BA's original authors. Without their contribution this paper would not exist. We communicated to them both our gratitude and the complete text, asking for feedback. We received three replies, the common denominator being that too much time has passed since the algorithm's conception in order for any substantial remarks to be made. One of the two main authors (the second one has not replied until the time of writing) emphasized that adjustments \textit{are} expected when applying mathematical ideas in practice.}
    \item an evaluation of 4 modern allocators across 8 workloads in terms of makespan, with respect to Robson's worst-case lower bound for general policies~\cite{robson74}
    \item a comparison, both in terms of makespan and page-local fragmentation, of allocator-generated placements against the respective BA solutions
\end{itemize}

The rest of this intellectual abstract is organized as follows: Section \ref{sec:fatality} elaborates on what led us tackle this work. Section \ref{sec:bck} provides the necessary background. Section \ref{sec:imp} describes our BA implementation. Section \ref{sec:res} includes the collected empirical evidence as well as a first discussion. Related works are listed in Section \ref{sec:rw}. Section \ref{sec:end} concludes our paper.

\section{Motivation and rationale}
\label{sec:fatality}
As mentioned, this paper is an immediate consequence of a sister work introducing 2DBP as a potentially useful tool for representing workload-allocator interaction. Despite it being infeasible to unpack everything done in that context, we try to summarize some key thoughts and link them to the work presented here.

\subsection{The need for a structured representation}
We build on the conjecture that \textit{returning to first principles is necessary} if a rigorous memory management theory is to be established~\cite{lamprakos2023viewing}. By first principles, fragmentation is the main enemy of any allocator, and it is a function of the interaction between workload behavior and allocator policy~\cite{wilson1995surv}.

To define fragmentation, the field employs functions of resident set size (RSS)~\cite{mesh}, this being probably an influence from the four alternative formulations proposed by Johnstone and Wilson in 1998~\cite{fragsolved}. No attempts have been made to evaluate each option's utility, in spite of the original authors encouraging such prospects. Whether general-purpose policies suffice to handle modern workloads or not is unclear, since works supporting both views exist~\cite{reconsider, learnalloc}.

Our sister work's research objective was to find a \textit{structured representation} capturing workload-allocator interaction, and a systematic approach which would enable this in practice for realistic workloads. We started by observing two distinct branches of DSA research: practical work intended to operate on realistic environments, and theoretical work exploring limits and other aspects of allocator policy.

Of particular interest to us was the resemblance of DSA to a variation of two-dimensional rectangular bin packing (2DBP)~\cite{chrobak1988some, buchsbaum}. Up to that point 2DBP had been treated as an NP-hard optimization problem~\cite{garey_jognson_1979}, with approximate algorithms generating placements of minimal makespan. We did not intend to create a novel 2DBP algorithm; optimizing virtual memory makespan is meaningless in the context of demand paging. But what if we viewed \textit{allocators themselves} as 2DBP ``algorithms'' with unknown optimization criteria? The resulting structures should multiplex enough of the workload-allocator interaction that we were targeting.

\begin{figure}[t!]%
    \centering
    \includegraphics[width=\columnwidth]{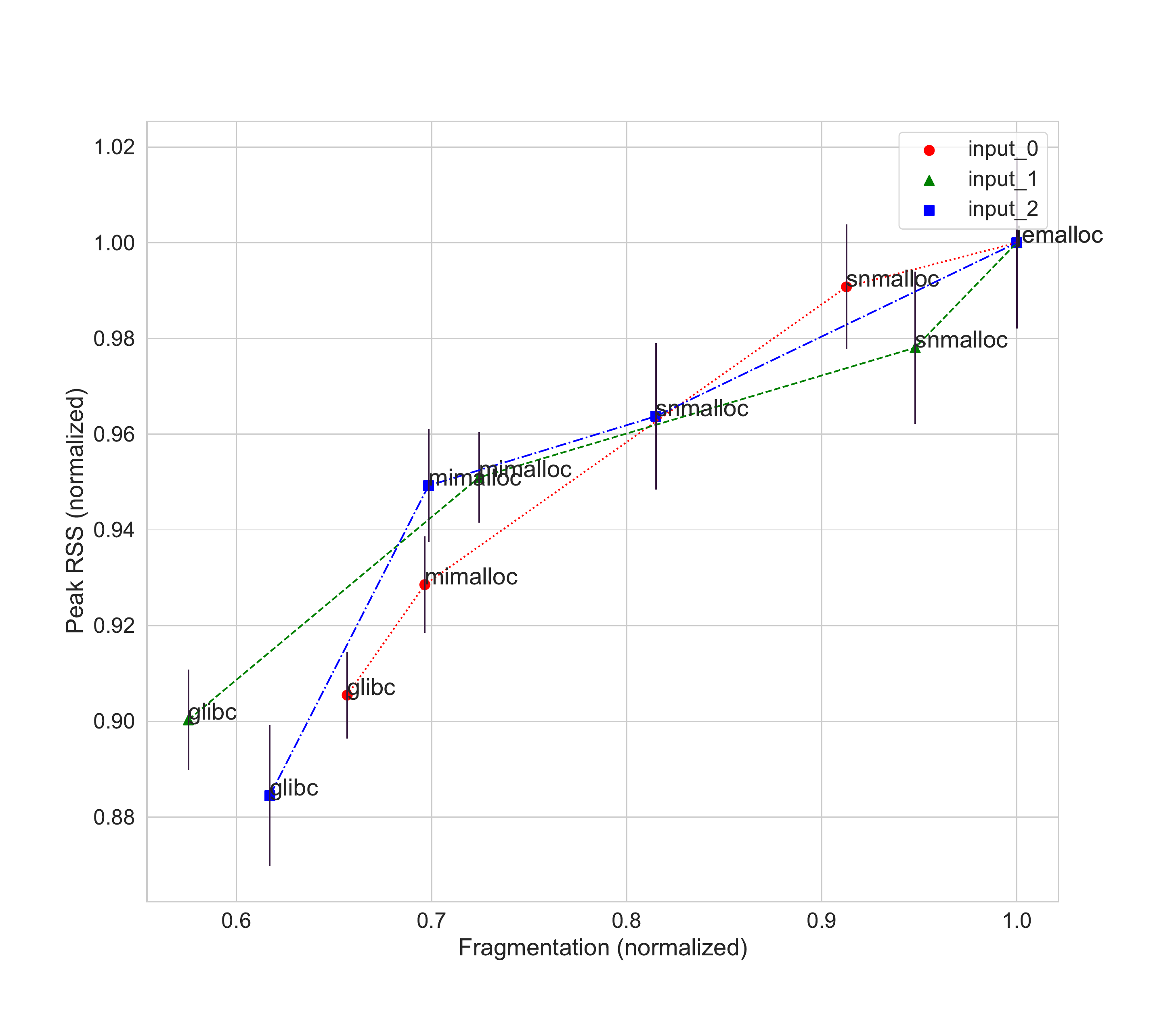}
    \caption{Scatter plot of three Linux \texttt{xmllint} workloads' peak RSS versus their 2DBP-based fragmentation across four modern allocators (\texttt{glibc}, \texttt{jemalloc}~\cite{jemalloc}, \texttt{mimalloc}~\cite{mimalloc} and \texttt{snmalloc}~\cite{snmalloc}. The black error bars are standard deviations of our RSS measurements. Fragmentation calculation is deterministic.}%
    \label{fig:motiv}%
\end{figure}

To this end, we devised a portable, trace-based simulation methodology for representing workload-allocator interaction as 2DBP instances. To conclude whether the representations produced bore any information of practical value, we investigated their relationship to maximum resident set size (RSS). We defined fragmentation in the 2DBP space, and measured it for 28 workloads linked to 4 modern allocators. For $46.4\%$ of the studied workloads, 2DBP-based fragmentation and maximum RSS exhibited a monotonic relationship as per Spearman's correlation coefficient ($\rho>0.65$). Lower fragmentation in 2DBP yielded up to $30\%$ smaller memory footprint in the real world. Figure \ref{fig:motiv} provides an example.

The fact that computations on trace-based simulation data correlated with empirical RSS measurements convinced us of 2DBP's potency. The non-uniformity in said correlation implies that, contrary to common practice, fragmentation is \textit{not} always the culprit behind RSS fluctuations.

\subsection{The logical conclusion of using 2DBP}
If representing workload-allocator pairs as 2DBP instances makes sense, then computing their approximately optimal counterparts and exploring how they relate could yield useful insights. For instance, until this point we do not possess any concrete idea on the bounds of real allocator placements, other than that real allocator placements on realistic inputs normally produce much better results than Robson's worst-case lower bounds~\cite{robson74, fragsolved, reconsider}. However, both Robson's bounds and offline 2DBP algorithms assume a single, contiguous mapping of memory--thus conceptualizing fragmentation as divergence from a placement's optimal makespan.

On the one hand, Linux virtual memory employing demand paging renders makespan irrelevant; allocators are commonly known to employ sparse addressing to combat fragmentation (this is what led us to define a page-local metric in the context of our sister work). On the other hand, Linux systems are not the only ones making use of dynamic memory allocation. In the embedded domain, statically pre-allocating and managing contiguous physical space is often normal. Following this line of thought to its logical conclusion, we get this intellectual abstract's rationale:

\begin{itemize}
    \item fragmentation is a context-dependent concept
    \item 2DBP is a context-free representation of fragmentation's source, that is, the interplay between program behavior and placement policy
    \item we do not know the achievable limits of allocators with respect to fragmentation
    \item we know 2DBP algorithms' achievable limits with respect to one specific fragmentation definition
    \item comparing 2DBP algorithms and allocators in 2DBP's common substrate could lead to fruitful results--a simple example being the ability to differentiate between programs that can, or cannot, benefit further from custom placement policies
\end{itemize}

\section{Background}
\label{sec:bck}
This section deals with the 2DBP formulation we are building on (Section \ref{sec:tdbp}) and the algorithm we are implementing (Section \ref{sec:buch}).
\subsection{Viewing Program-Allocator Interaction as a Bin Packing Instance}
\label{sec:tdbp}
Assume that an application is executed and, as regards dynamic memory, it is served by some specified allocator. Along its course the application will have generated a sequence of $K$ requests $R = \{R_0, R_1, ..., R_{K-1}\}$. For simplicity, assume that for each request $R_i, i \in \{0, ..., {K-1}\}$ the following is true:

\begin{equation}
    R_i = 
        \begin{cases}
            M(n) & \text{(allocate n bytes)}\\
            F(j), \; j < i & \text{(free memory allocated for request $R_j$)}
        \end{cases}
\end{equation}

An allocator can be considered as a function $A()$ operating on request sequences like $R$. Then the allocator's output after processing the last request is a set of $N=K/2$ \textit{placed} memory blocks, which from now on we will refer to as ``jobs'':

\begin{equation}
    \label{eq:apl}
    A(R) = J_{{PA}} = \{J_{PA,0}, ..., J_{PA,{N-1}}\}
\end{equation}
\begin{equation}
    \label{eq:pl}
    J_{PA,i} = (t_{sA, i}, t_{eA, i}, h_{A, i}, p_{A, i})
\end{equation}

The subscript $PA$ means ``placed by allocator $A$'', $i$ is a unique job identifier ($i \in \{0, ..., {N-1}\}$), $t_{sA, i}$ is the point in time when job $i$ was allocated, $t_{eA, i}$ the respective time of deallocation, $h_{A, i}$ is the size of the memory block that $A$ allocated (different allocators spawn different-sized blocks for same-sized requests, according to their policy with respect to size classes and block metadata), and $p_{A, i}$ is the virtual address where the job was placed. Time is measured in allocated bytes, and progresses forward based on the rule below:

\begin{equation}
    t(R_i) =
        \begin{cases}
            0 & \text{initially}\\
            t(R_{i-1}) & \text{iff } R_i = F(j): \; j < i\\
            t(R_{i-1}) + h_{A, i} & \text{iff } R_i = M(n)
        \end{cases}
\end{equation}

\begin{figure}[t!]%
    \centering
    \includegraphics[width=\columnwidth]{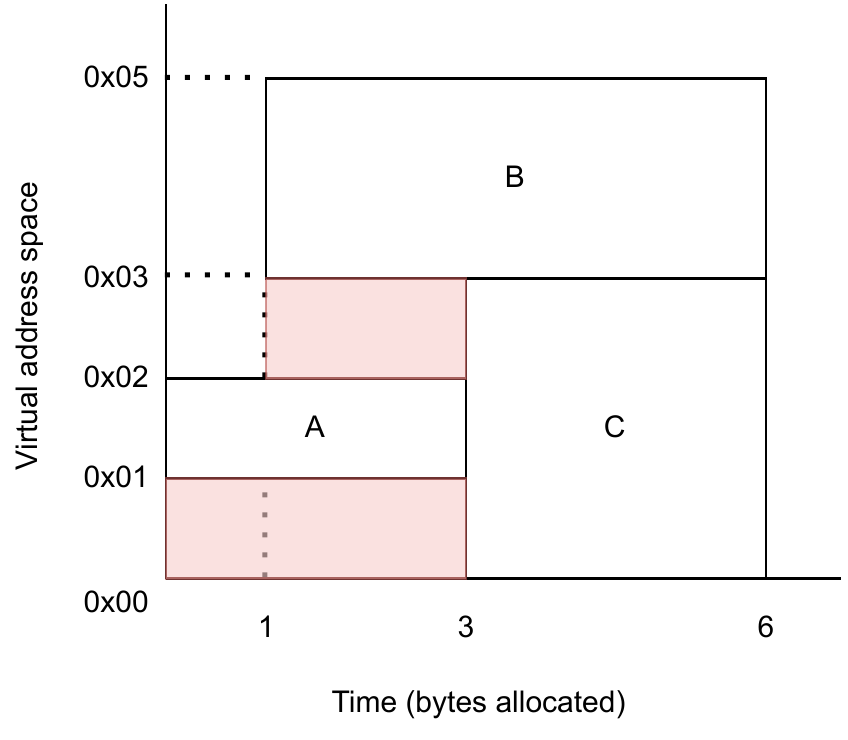}
    \caption{A simple 2DBP example. Consider the following requests sequence: (i) \texttt{A = malloc(1)}, (ii) \texttt{B = malloc(2)}, (iii) \texttt{free(A)}, (iv) \texttt{C = malloc(3)}, (v) \texttt{free(B)} and (vi) \texttt{free(C)}. This figure combines the aforementioned requests with an imaginary allocator's responses, placing block A at virtual address 0x01, block B at 0x03 and block C at 0x00. The horizontal axis measures time in allocated bytes. Time progresses forward after each allocation request, and remains unaltered after each deallocation request. Our sister work's proposal regarding fragmentation is indicated by the two shaded rectangles. They represent segments which the allocator left unused, thus reserving higher addresses in order to handle all requests.}
    \label{fig:frag}%
\end{figure}

Let us also assume the below statements hold:

\begin{itemize}
    \item the served application is \textit{single-threaded} and \textit{deterministic}. Every time it is executed with the same input, it produces exactly the same request sequence
    \item the allocator places all jobs in the \textit{same} memory mapping\footnote{By ``memory mapping'' we refer to contiguous virtual address space managed by the allocator--as obtained via \href{https://linux.die.net/man/2/sbrk}{\texttt{sbrk}} or \href{https://linux.die.net/man/2/mmap}{\texttt{mmap}}.}
    \item there are no memory leaks, no double frees, and more generally the requests sequence is \textit{well-formed}
\end{itemize}

The jobs sequence $J_{PA}$ can be viewed as the solution that allocator $A$ devised for the two-dimensional bin packing (2DBP) problem defined by a corresponding sequence of \textit{unplaced} jobs $J$:

\begin{equation}
    \label{eq:aupl}
    J = \{J_{0}, ..., J_{N-1}\}
\end{equation}
\begin{equation}
    \label{eq:upl}
    J_{i} = (t_{s, i}, t_{e, i}, h_{i})
\end{equation}

However, instead of optimizing for the final placement's makespan as normally happens in 2DBP, allocator $A$ placed each job in $J$ according to some unknown criterion implied by its (also unknown) policy. 

We now note down some more definitions that will be of use later. First, a job is considered \textit{live} in the open interval $(t_{s, i}, t_{e, i})$. Thus we define the \textit{liveness function} $a(J_i, t)$:

\begin{equation}
    \label{eq:live}
    a(J_i, t) =
        \begin{cases}
            1 & t_{s, i} < t < t_{e, i} \\
            0 & \text{elsewhere}
        \end{cases}
\end{equation}

The \textit{load} at some particular moment $t$ corresponds to the sum of heights belonging to jobs that are alive at $t$:

\begin{equation}
    \label{eq:load}
    l(t) = \sum_{i = 0}^{N-1} {a(J_i, t)h_i}
\end{equation}

Across its lifetime, a job contributes an \textit{individual load} $|J_i| = (t_{e,i} - t_{s, i})h_i$. A series of jobs $J$ is characterized by its \textit{total load} $L_T=\sum_{i=0}^{N-1}{|J_i|}$. It is also characterized by its \textit{maximum load} $L=l(t_{tL}):l(t_i) \leq L, \; \forall i \in \{0, ..., N-1\}$.

A fitting analogy for a 2DBP instance is a Tetris game: we want to minimize the gaps between placed blocks. The fragmentation metric we propose is the gaps-to-total-load ratio. Returning to the allocator placement in Eq. \ref{eq:apl}, we traverse it and record all gaps between jobs belonging \textit{to the same virtual page}. Recall that in the context of virtual memory with demand paging treating gaps between pages as sources of fragmentation is meaningless. 

A placement $J_{PA}$ thus implies, beyond its jobs, a second sequence of rectangles $G$, corresponding to gaps between jobs as defined in this paragraph. Assuming that the number of those gaps is $N_G$, fragmentation can be quantified as:

\begin{equation}
    \label{eq:frag}
    F = \frac{\sum_{j=0}^{N_G} {|G_j|}}{L_T}
\end{equation}

\begin{figure}[t!]%
    \centering
    \includegraphics[width=\columnwidth]{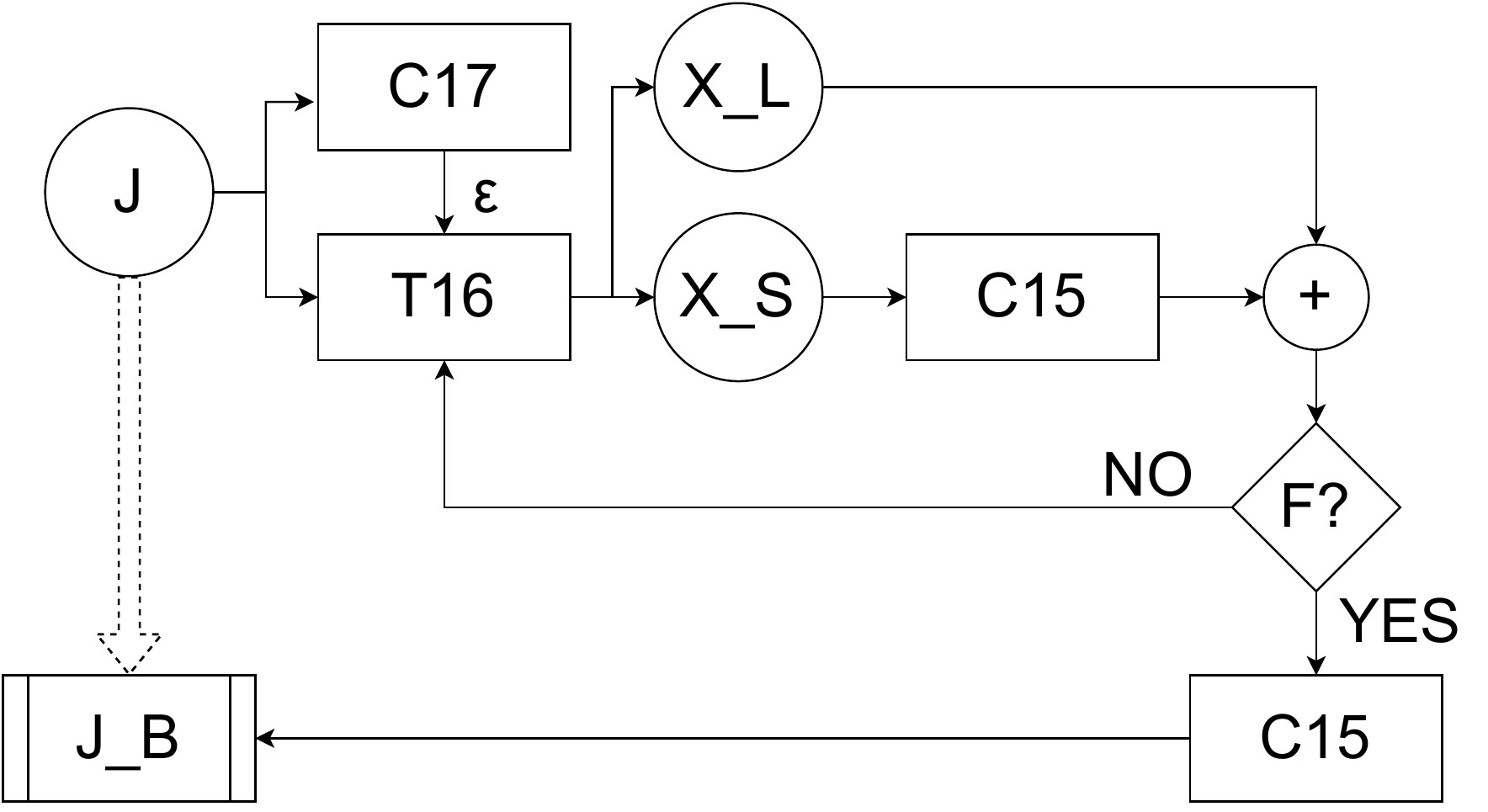}
    \caption{BA \textit{in theory.} $J$, $X_L$ and $X_S$ denote intermediate sets of boxes. The $+$ operator merges sets of boxes together in a single set. The $F$ condition is explained in the text. The three rectangles are BA's basic pillars: Corollary 17 (Section \ref{sec:c17}), Theorem 16 (Section \ref{sec:t16}), and Corollary 15 (Section \ref{sec:c15}). $J_B$ is the algorithm's end product, that is a set of boxes of \textit{identical height}. The double dotted arrow indicates how the initial sequence of different-sized jobs is transformed, after BA's application, into a set of same-sized boxes.}
    \label{fig:happy}
\end{figure}

\subsection{The Algorithm}
\label{sec:buch}
BA operates on inputs of the form defined in Eqs. \ref{eq:aupl} and \ref{eq:upl}. It produces, as the abstract allocator in Eqs. \ref{eq:apl} and \ref{eq:pl}, a series of placed jobs. At each point in time, addresses are occupied by at most one job. 

The criterion optimized is the output placement's makespan $M$, that is the maximum address used for a placement. BA produces approximately optimal solutions. In particular, BA guarantees that $M \leq [1 + O((h_{max}/L)^{1/7})]L$. The same paper describes a stronger flavor producing placements with $M \leq (2+\epsilon)L$ for every $\epsilon > 0$.\footnote{If the strength relationship between the two algorithms is not evident, consider the following cases: 

(i) for $h_{max} = L$, the weak algorithm guarantees that $M \leq [1+O(1)]L \Rightarrow \exists \; c_1 > 0: \; M \leq (1+c_1)L$. To outperform the strong version, it should hold that $1+c_1 < 2 + \epsilon \; \Rightarrow c_1 < 1 + \epsilon$. Arguably such instances may appear, but since we know nothing about the range of $c_1$, we must reason according to the worst case--and assign much higher probability to $c_1$ exceeding $1 + \epsilon$.

(ii) for $h_{max} = L/128$, we similarly arrive to the condition $c_2 < 2(1+\epsilon)$. Recall that the strong algorithm works for all $\epsilon > 0$, so we may as well consider it small enough to require $c_2 < 2$. Again, we have no evidence suggesting such a tight range for $c_2$.}

We managed to implement only the weaker version of BA, for reasons that will be explained later (Section \ref{sec:t19}). We expand on the weaker one for now. Its overview is shown at Figure \ref{fig:happy}. Regarding individual components, i.e., theorems and corollaries derived in the original paper, we follow the naming and numbering established in the STOC proceedings version~\cite{buchsbaum}. Having a copy of it available side by side with this intellectual abstract is more than advised, if possible. 

We have tried to make the present text self-contained to the greatest degree, in the sense of depicting and elaborating on everything that we implemented from the original. Still, BA is understandably complex. Apart from consulting the original publication, we also suggest reading the current section's contents \textit{twice}: once in their default order and once more in reverse.

\begin{figure}[t!]%
    \centering
    \includegraphics[width=\columnwidth]{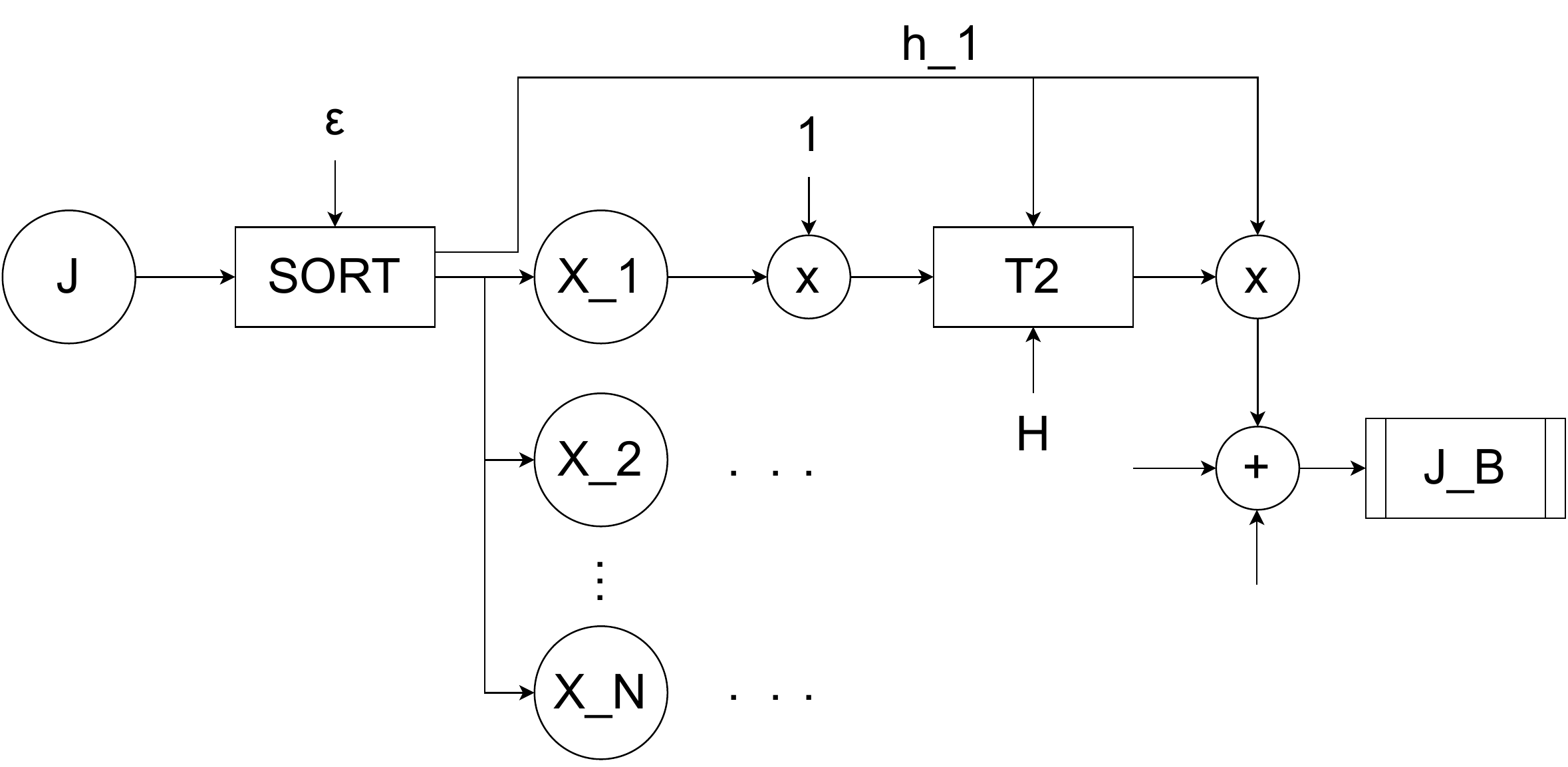}
    \caption{Corollary 15, unpacked. A new component, that is Theorem 2, appears here. This will be unpacked later on. A new operator $x$ is also shown. It is used to change a job's height. For example it changes the jobs in $X_1$, initially of height $h_1$, to jobs of unit height.}
    \label{fig:c15}
\end{figure}

\subsubsection{Main idea} 2DBP is NP-hard due to the variety of sizes present in sequence $J$; if all rectangles had the same height, the problem's optimal solution could be derived via interval graph coloring (IGC). BA exploits this fact.

BA views all data as \textit{boxes}, defined by triplets identical to Eq. \ref{eq:upl}. Jobs in $J$ are boxes containing nothing. New boxes, containing existing ones, are created in multiple parts of the algorithm. The ultimate goal is to \textit{box all jobs in $J$ into a series of boxes of identical height.} Then IGC can be applied to find an optimal placement. Figure \ref{fig:happy} shows how $J$ is transformed to a set $J_B$ comprising same-sized boxes. From this point onward, whenever the term ``job'' is used, we could either refer to actual jobs or boxes. For BA \textit{everything is a box}.

\subsubsection{Corollary 17}
\label{sec:c17}
This is the algorithm's entry point.\footnote{Treating corollaries and theorems as if they were execution stages may seem weird. But every proof in the original paper is given by construction; a claim is made, then the operations to apply to the input are described in terms of earlier corollaries and theorems, and finally the claim is validated based on properties of the involved operations. Thus following BA in its entirety amounts to following how each component is applied to the rest.} Its only function is to calculate the error parameter $\epsilon$, to be given as input to Theorem 16. $\epsilon$ is defined as ${(h_{max}/L)}^{1/7}$. An intuition-friendly way to view it is that the bigger it is, the less optimal will the overall boxing be.

\subsubsection{Theorem 16}
\label{sec:t16}
This stage operates on arbitrary sets of jobs, and expects an error parameter $\epsilon \in (0, 1]$ as additional input. Then the elements $r=h_{max}/h_{min}$, $\mu=\epsilon/{\log^2 r}$ and $H=\lceil {\mu^5 h_{max} / \log^2 r} \rceil$ are computed. Then jobs are divided in two disjoint subsets:

\begin{itemize}
    \item $X_S$: jobs of height at most $\mu H$
    \item $X_L$: the rest of the jobs
\end{itemize}

Corollary 15 is applied to $X_S$ with error parameter $\epsilon = \mu$ and height parameter $H$. This yields a set of boxes of height $H$. The boxes are merged with the jobs of $X_L$, $r$ and $\mu$ are recomputed, and then one of two possible scenarios holds; either $\log^2 r \geq 1/\epsilon$ or $\log^2 r < 1/\epsilon$. In the first case, Theorem 16 is recursively applied to the merged set with error parameter $\epsilon$. Otherwise, a last application of Corollary 15 is made. On Figure \ref{fig:happy} this condition check is represented as $F$.

\subsubsection{Corollary 15}
\label{sec:c15}
This stage accepts (i) a set of jobs $J$, (ii) an error parameter $\epsilon > 0$ and (iii) a height parameter $H$. Jobs in $J$ must be of height at most $\epsilon H$; this holds by default in Figure \ref{fig:happy} due to the definition of $X_S$.

Figure \ref{fig:c15} illustrates this stage: jobs are sorted in buckets according to the inequality ${(1 + \epsilon)}^{i-1} < h \leq {(1 + \epsilon)}^{i}$. Jobs belonging to the same bucket are rounded up so that their height is $h_i = \lfloor {(1 + \epsilon)}^{i} \rfloor$. Thus in the figure, all the jobs in $X_1$ have height $h_1$, etc. Each bucket is then downscaled to unit height and Theorem 2 is called with height parameter $\lfloor H/h_i \rfloor$. Theorem 2's output is upscaled to boxes of height $H$ and all box subsets, i.e. as created by a parallel application of the described flow to every $X_i$, are merged to a single set $J_B$ comprising boxes of height $H$.

\begin{figure}[t!]%
    \centering
    \includegraphics[width=\columnwidth]{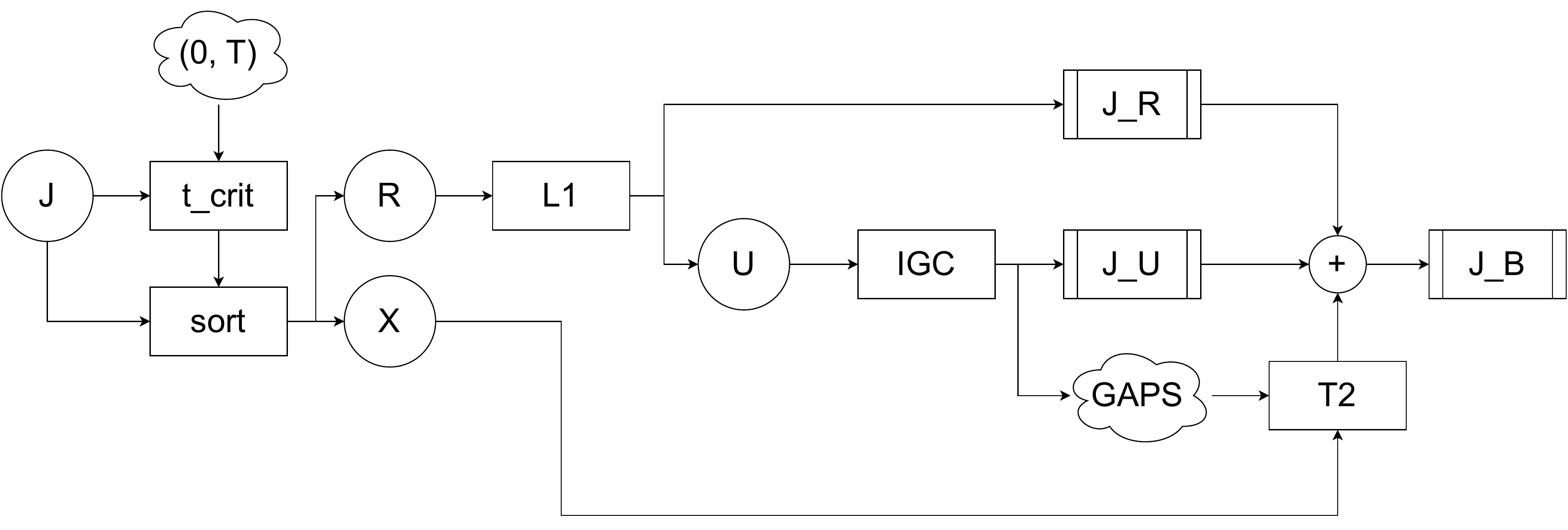}
    \caption{Theorem 2, unpacked. Lemma 1 and interval graph coloring (IGC) also appear.}
    \label{fig:t2}
\end{figure}

\subsubsection{Theorem 2}
\label{sec:t2}
This is the most complex part of BA, and is a recursive procedure. It is depicted at Figure \ref{fig:t2}. Theorem 2 expects as inputs (i) a set of jobs of unit height and (ii) a set of \textit{bounding intervals} represented as clouds in the figure. Initially the bounding interval equals to the whole time horizon of the jobs in $J$, plus a random moment at which at least one job is live. Bounding intervals correspond to temporal segments used to sort jobs in $J$, this time not according to their heights but their liveness:

\begin{itemize}
    \item if we name all the endpoints of all the bounding intervals as \textit{critical times}, $R$ contains jobs that are live at least at one critical time
    \item set $X$ holds jobs that are not live at any critical time
\end{itemize}

Lemma 1 is then applied to the jobs in $R$. As a result, a set of boxed jobs $J_R$ and a set of unresolved jobs $U$ are created. The jobs in U are packed via IGC, and from that packing a new set of bounding intervals is created besides $J_U$. These feed a deeper invocation of Theorem 2, which focuses on the jobs in $X$. At the last call of Theorem 2, $X$ is expected to be empty. Boxed jobs are returned and consolidated with products of shallower recursive layers.

\subsubsection{Lemma 1}
\label{sec:l1}
This is BA's cornerstone, and straightforward enough not to require an illustration. A set of jobs of unit height, all live at a specific moment $t$ and a height parameter $H$ are the inputs. A set of boxes $J_B$ and a set of unresolved, i.e. unboxed jobs $U$ are the outputs.

We omit box derivation since it mostly serves the original publication's mathematical arguments, leading to the makespan-related guarantees of the overall algorithm. Elaborating further on its internals is outside this paper's scope.

\begin{figure}[t!]%
    \centering
    \includegraphics[width=\columnwidth]{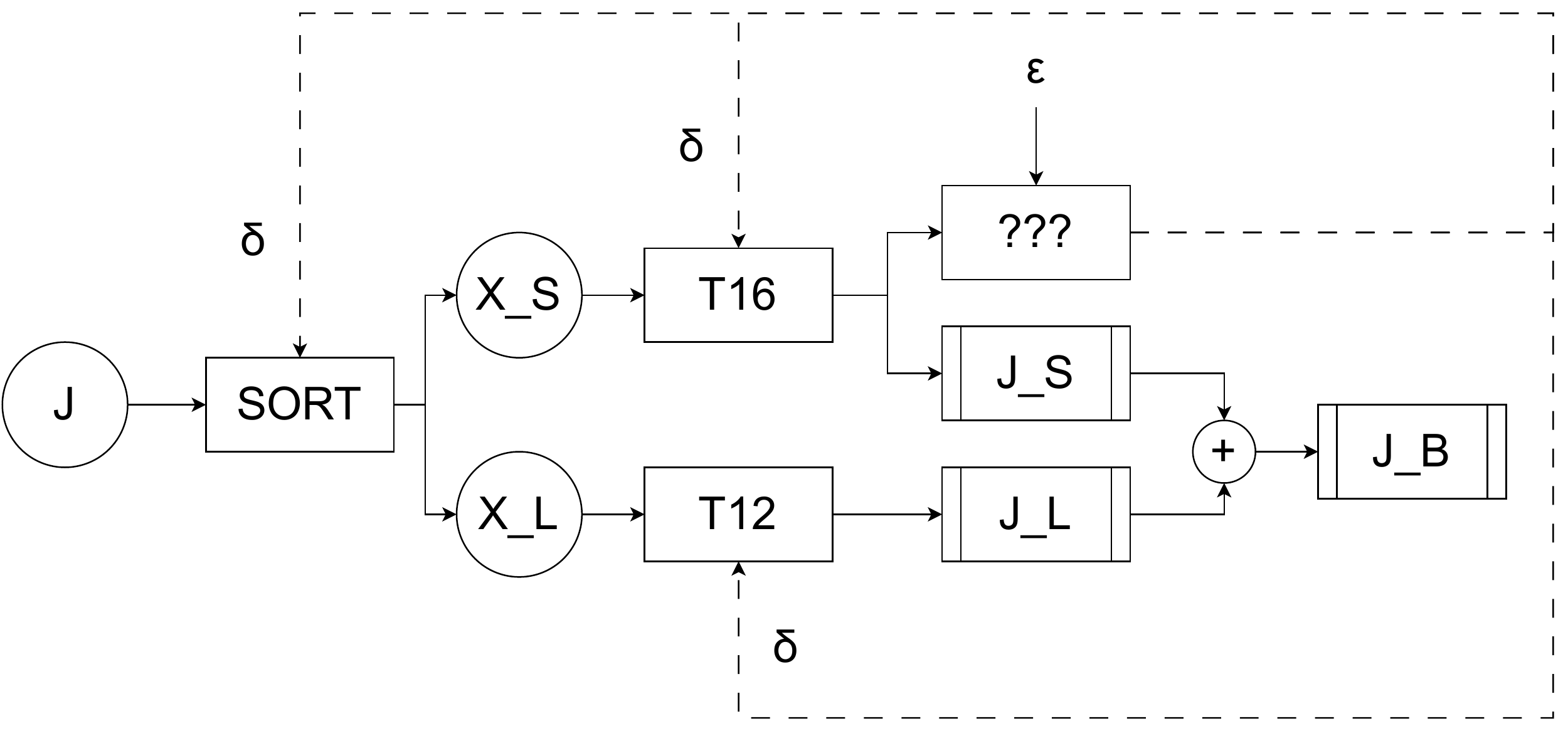}
    \caption{Theorem 19. Continuous arrows denote normal progression of time. Dashed arrows are time-travelling information. We did not succeed in implementing this.}
    \label{fig:t19}
\end{figure}

\subsubsection{Theorem 19}
\label{sec:t19}
As mentioned in Section \ref{sec:buch} there is a stronger algorithm in the BA paper than Corollary 17. It is shown in Figure \ref{fig:t19}. Its inputs are an arbitrary set of jobs and an error parameter $\epsilon > 0$.

The main difficulty posed by this theorem is that it uses information from the future. It commences with a height-based sorting operation according to some ``small positive $\delta$'' that has not yet been computed. It then applies Theorem 16 to the first subset, yielding an $(1 + c\delta)$-approximation. \textit{Then} it computes $\delta$ via the equation:
\begin{equation}
\label{eq:magic}
\delta(c+1) = \epsilon    
\end{equation}
This information must travel back in time to feed the sorting operation and Theorem 16 itself! We emphasize this difficulty with the ``???'' component in Figure \ref{fig:t19}.

A workaround we thought was some form of speculative execution: choose $\delta$ at random, and if it satisfies Eq. \ref{eq:magic} within some acceptable error range, proceed with the remaining steps, else retry. We have not explored this idea further.

\section{Implementation}
\label{sec:imp}
Let us turn to our implementation. This exposition does not intend to be of software engineering character; we will not allocate space to describe the \textit{particular} tools used, e.g., programming language, sorting algorithms, jobs representation etc. We shall focus on the high-level obstacles found and the high-level mechanisms devised to overcome them. More seasoned developers will come up with more efficient mechanisms, but they will stumble on the same intricacies that we now come to discuss.

\begin{table}[H]
    \centering
    \caption{An example of what led us to adjust BA. Theorem 16 falls in an infinite loop if the height threshold defining $X_L$ and $X_S$ is smaller than the smallest height in $J$.}
    \label{tab:motiv}
    \begin{tabular}{|c|c|c|c|}
        \hline
        \textbf{$h_{min}$} & \textbf{$h_{max}$} & \textbf{$(\epsilon_t,h_t)$} & \textbf{$(\epsilon_p,h_p)$} \\
        \hline
        8 & 524288 & (0.76, 0.033) & (6.19, \hspace{5pt}8.27) \\ \hline
        8 & 1048576 & (0.97, 0.037) & (6.55, 10.01) \\ \hline
        16 & 524288 & (0.75, 0.037) & (6.12, 18.91) \\ \hline
        8 & 75497472 & (0.98, 0.020) & (6.59, \hspace{5pt}9.62) \\
        \hline
    \end{tabular}
\end{table}

\begin{figure*}[t!]%
    \centering
    \includegraphics[width=\textwidth]{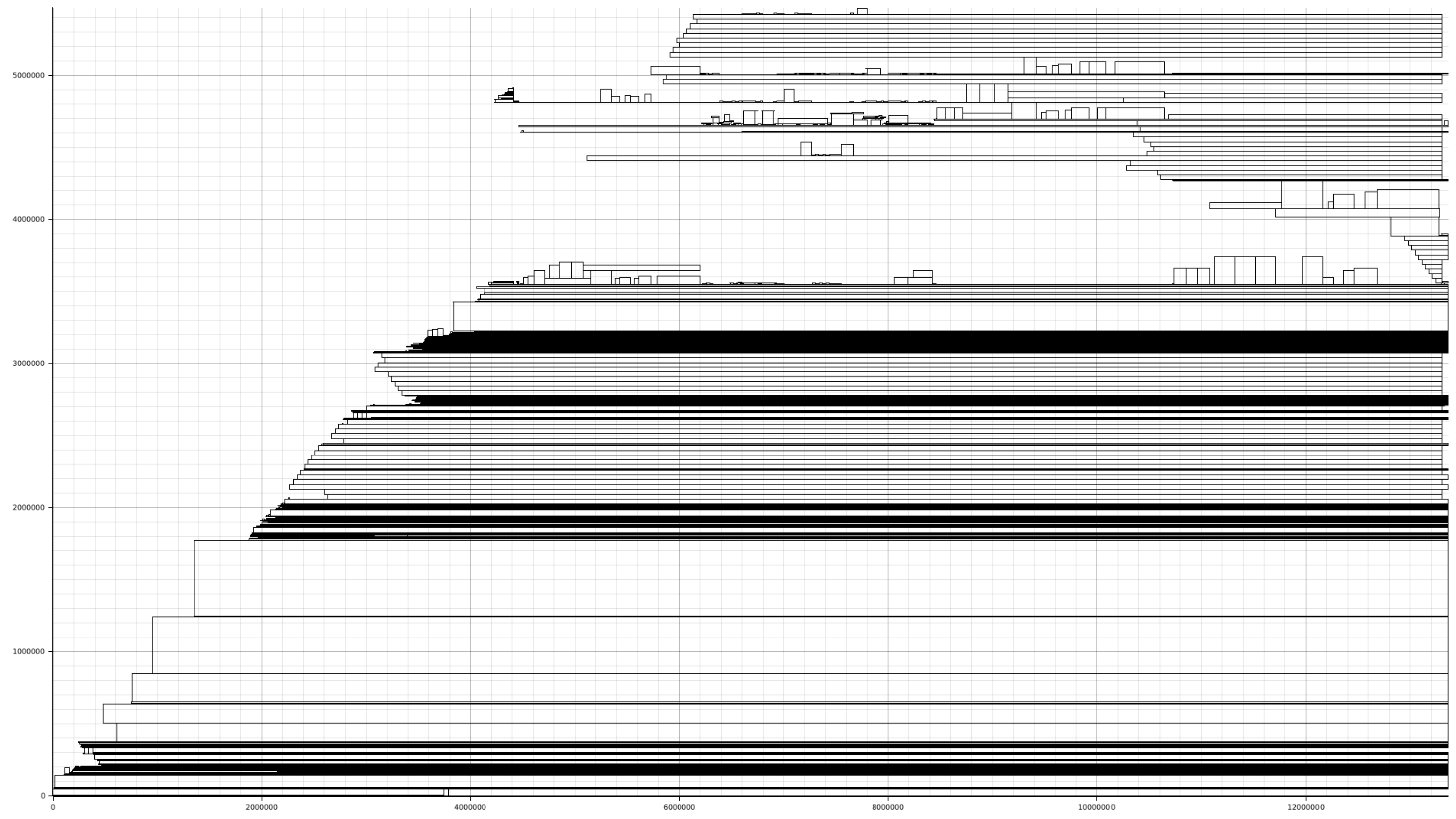}
    \caption{A sample placement for the \texttt{bork} benchmark, produced by our BA implementation.}%
    \label{fig:temple}%
\end{figure*}

An example is shown in Table \ref{tab:motiv}. The first two columns show the characteristics of some indicative workloads. The third column contains (i) the theoretical $\epsilon$ that Corollary 17 computed and (ii) the corresponding threshold used by Theorem 16 to split its jobs in $X_S$ and $X_L$. Given that Theorem 16 is recursive, if at some point $X_S$ is empty, the algorithm falls in an infinite loop. This would have happened for virtually every input if we had relied on the theoretical $\epsilon$ provided by Corollary 17. The fourth column of Table \ref{tab:motiv} shows the corresponding values that allowed Theorem 16 to converge. In that case, the practical $\epsilon$ values used are well outside the $(0, 1]$ range that BA in theory demands.

At this point we could either discontinue our effort or devise a fault-tolerant scheme that would allow us to proceed. This could not be done for each component in isolation, since it must be evident by now that there are strong dependencies from each stage to the next. We chose the second option. Our design was based on two key observations:

\begin{itemize}
    \item components can be broken into \textit{safe} and \textit{unsafe} sections. For instance, merging distinct sets of boxes to a single set is safe. Calling another component is unsafe
    \item most of the time, the failure of unsafe sections is owed to the $\epsilon$ value computed by Corollary 17
\end{itemize}

The next sections explain how we put these observations to good use.

\subsection{Controlling unsafe parts with failure propagation}
BA has a cascade structure. After the error parameter $\epsilon$ has been computed, jobs pass through the other components getting re-boxed at each stage. If we view each theorem and corollary as a separate function, the stack trace starts from Corollary 17 and grows across the rest of the pipeline.

\begin{figure}[H]%
    \centering
    \includegraphics[width=\columnwidth]{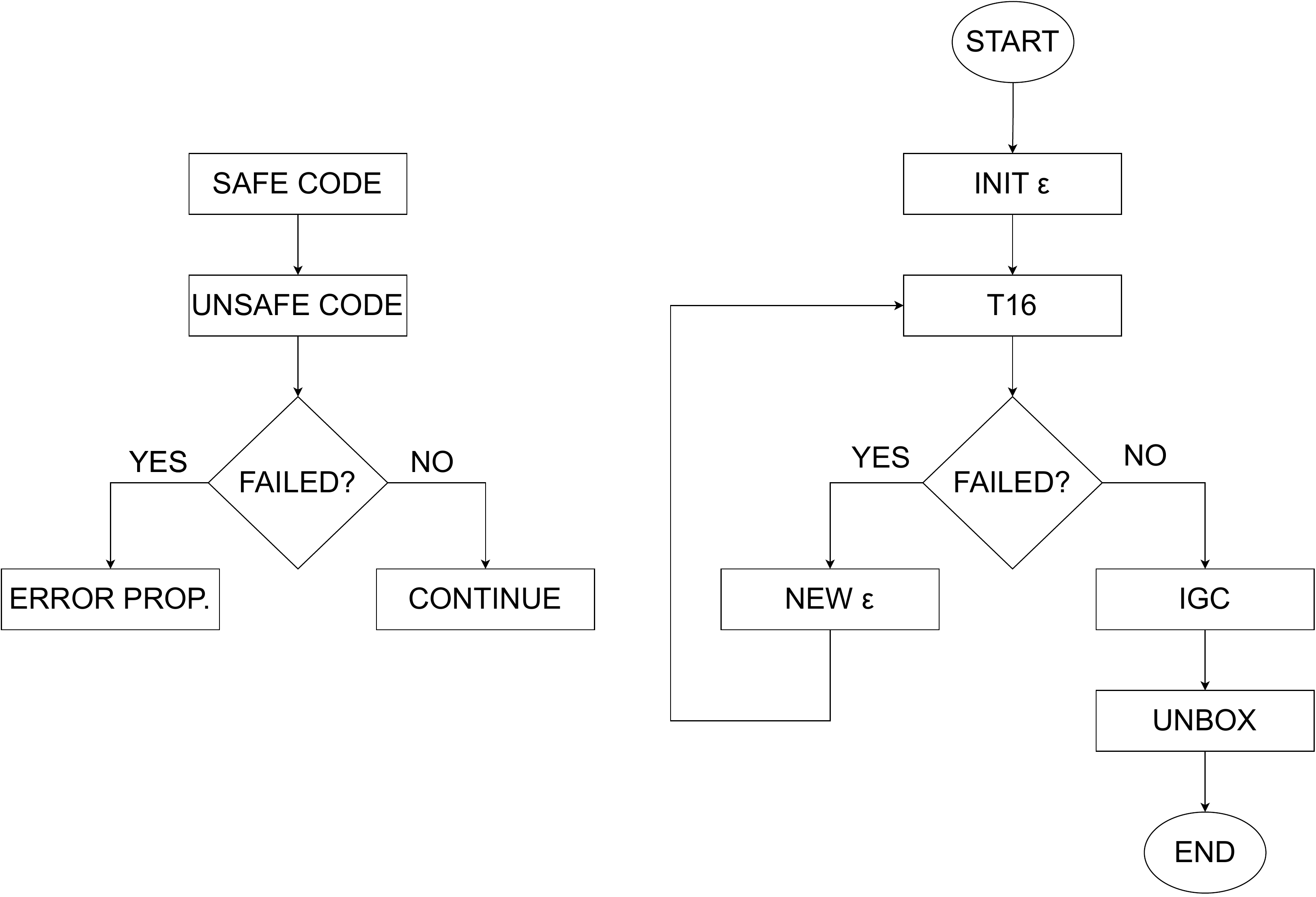}
    \caption{Our fault-tolerance mechanism divides each BA component to safe and unsafe portions. Each unsafe portion has been mapped to a specific criterion that decides whether the portion failed or not. At failure an error signal is propagated back across the pipeline, ending in BA's entry point, i.e., Corollary 17. This has been adjusted in order to change the value of $\epsilon$ upon failure detection (right).}
    \label{fig:safe}
\end{figure}

By trial and error, we deduced that $\epsilon$ is most often the culprit for some component not working as expected. Our idea is thus to propagate, upon detection of failure, a signal towards the root of the stack trace--that is to BA's entry point, Corollary 17. As shown in the right part of Figure \ref{fig:safe}, on such occasions a new $\epsilon$ is computed and the overall flow is retried. A large amount of effort was thus invested in identifying component sections that may lead to failure (``unsafe code'' in Figure \ref{fig:safe}). An example has already been given via Table \ref{tab:motiv}. 

Once the identification process was complete, implementing a fault-tolerance mechanism was trivial. We just checked a case-specific condition immediately after executing unsafe code. If the condition's result was ``error'', we propagated an error signal back to the caller. All calls to component functions are considered unsafe points on their own: consequently, the caller checks its own error condition upon the callee's return, and if an error is found it is sent further back until it reaches Corollary 17, which adjusts $\epsilon$.

\begin{figure*}[t!]%
    \centering
    \includegraphics[width=\textwidth]{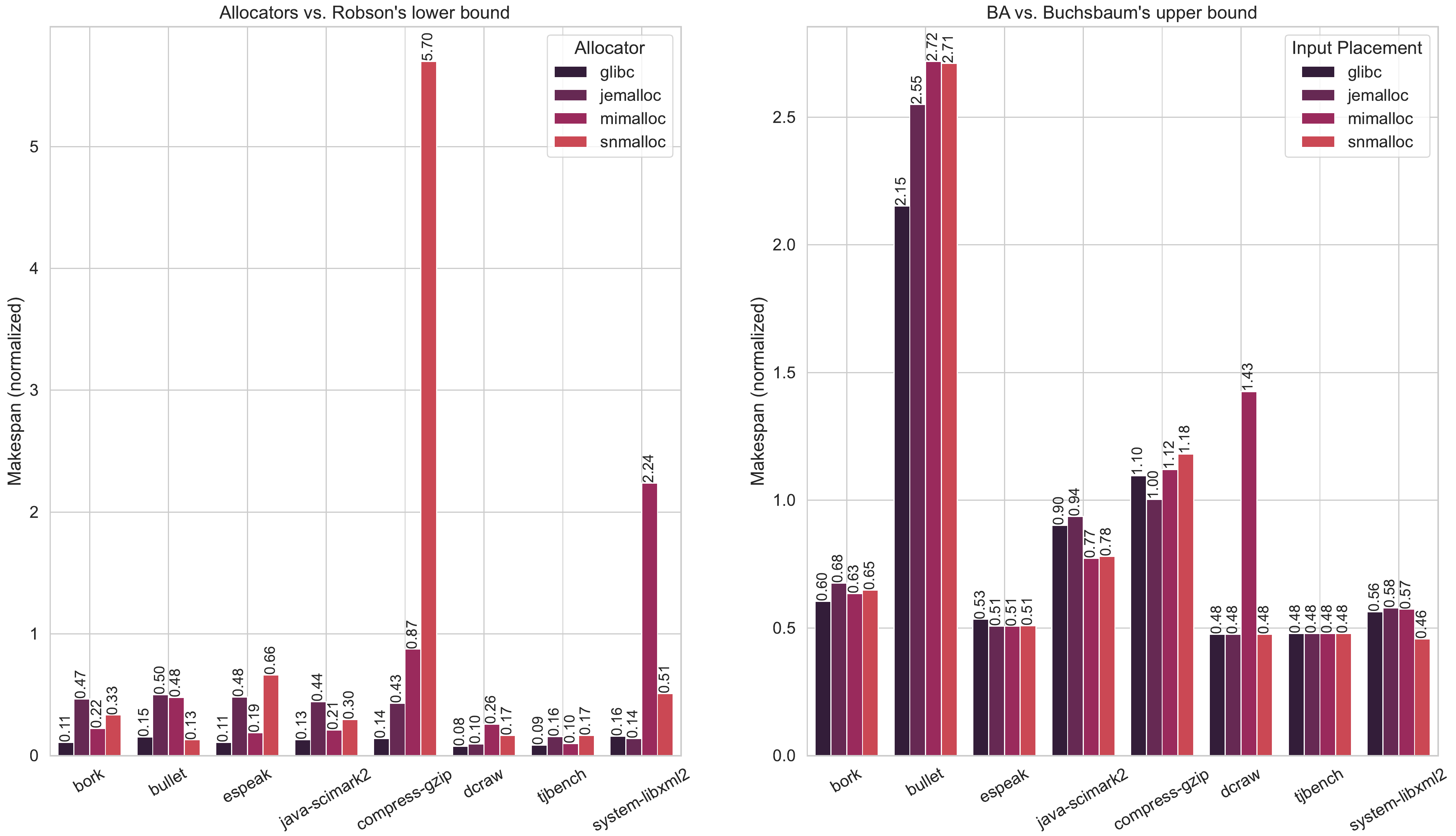}
    \caption{Left: allocators' makespan normalized to Robson's worst-case lower bound ($\frac{1}{2}L\cdot log_2{h_{max}}$). Right: BA's makespan normalized to its corresponding upper bound, $[1 + 2\cdot (h_{max}/L)^{1/7}]L$. Note that we replaced the original bound's big-O notation with a factor of 2.}%
    \label{fig:bounds}%
\end{figure*}

\begin{figure*}[t!]%
    \centering
    \includegraphics[width=\textwidth]{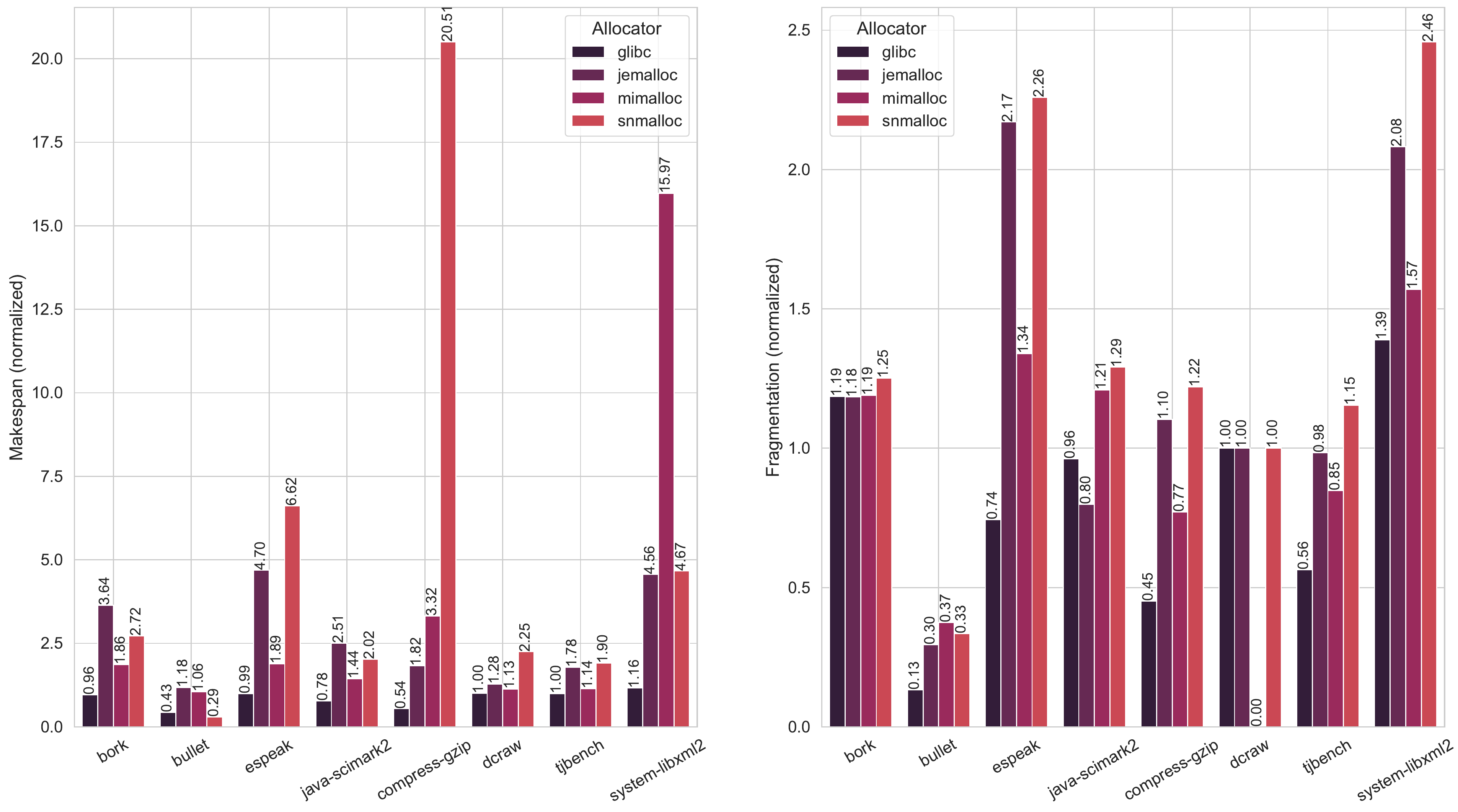}
    \caption{Left: allocators' makespan normalized to the corresponding BA placement's makespan. Right: the analogous data for page-local 2DBP-based fragmentation. Particularly about \texttt{dcraw} and \texttt{mimalloc}, the allocator scored a perfect 0 fragmentation while BA did worse--hence the zero value in the figure.}%
    \label{fig:versus}%
\end{figure*}

\begin{table*}
    \caption{The complete set of measurements we took in order to form the experimental results discussed. The only extra information here is the number of jobs per workload. The present data are continued in Table \ref{tab:frag2}.}
    \label{tab:frag1}
    \centering
    \begin{tabular}{|p{0.1\textwidth}|l|c|c|c|c|l|c|c|}
        \hline
        \textbf{Workload (\#jobs)} & \textbf{Allocator} & $\mathbf{L}$ & $\mathbf{h_{max}}$ & \multicolumn{1}{p{0.03\textwidth}|}{\textbf{Robson bound}} & \multicolumn{1}{p{0.03\textwidth}|}{\textbf{BA bound}} & \textbf{Setting} & \textbf{Makespan} & \textbf{Fragmentation} \\ \hline
        \multirow{8}{0.1\textwidth}{bork (12837)} & \multirow{2}{*}{glibc} & \multirow{2}{*}{4919904} & \multirow{2}{*}{528376} & \multirow{2}{*}{46766652} & \multirow{2}{*}{8869297} & idealloc & 5365096 & 0.149 \\ \cline{7-9}
        & & & & & & real & 5130728 & 0.176 \\ \cline {2-9}
        & \multirow{2}{*}{jemalloc} & \multirow{2}{*}{4923384} & \multirow{2}{*}{524288} & \multirow{2}{*}{46772148} & \multirow{2}{*}{8870789} & idealloc & 5993192 & 0.147 \\ \cline{7-9}
        & & & & & & real & 21826256 & 0.174 \\ \cline {2-9}
        & \multirow{2}{*}{mimalloc} & \multirow{2}{*}{4956152} & \multirow{2}{*}{524288} & \multirow{2}{*}{47083444} & \multirow{2}{*}{8926065} & idealloc & 5664904 & 0.150 \\ \cline{7-9}
        & & & & & & real & 10550912 & 0.178 \\ \cline {2-9}
        & \multirow{2}{*}{snmalloc} & \multirow{2}{*}{5267888} & \multirow{2}{*}{524288} & \multirow{2}{*}{50044936} & \multirow{2}{*}{9450893} & idealloc & 6132336 & 0.151 \\ \cline{7-9}
        & & & & & & real & 16695280 & 0.189 \\ \hline
        \multirow{8}{0.1\textwidth}{bullet (79950)} & \multirow{2}{*}{glibc} & \multirow{2}{*}{35388688} & \multirow{2}{*}{26742776} & \multirow{2}{*}{436566283} & \multirow{2}{*}{72928235} & idealloc & 156912368 & 0.256 \\ \cline{7-9}
        & & & & & & real & 67285248 & 0.034 \\ \cline {2-9}
        & \multirow{2}{*}{jemalloc} & \multirow{2}{*}{38789376} & \multirow{2}{*}{29360128} & \multirow{2}{*}{481130908} & \multirow{2}{*}{79945817} & idealloc & 203864400 & 0.047 \\ \cline{7-9}
        & & & & & & real & 240987280 & 0.014 \\ \cline {2-9}
        & \multirow{2}{*}{mimalloc} & \multirow{2}{*}{38101248} & \multirow{2}{*}{27262976} & \multirow{2}{*}{470558789} & \multirow{2}{*}{78204253} & idealloc & 212559984 & 0.043 \\ \cline{7-9}
        & & & & & & real & 224395264 & 0.016 \\ \cline {2-9}
        & \multirow{2}{*}{snmalloc} & \multirow{2}{*}{45991616} & \multirow{2}{*}{33554432} & \multirow{2}{*}{574895200} & \multirow{2}{*}{94534066} & idealloc & 256191312 & 0.042 \\ \cline{7-9}
        & & & & & & real & 74727936 & 0.014 \\ \hline
        \multirow{8}{*}{espeak (984)} & \multirow{2}{*}{glibc} & \multirow{2}{*}{1549032} & \multirow{2}{*}{675832} & \multirow{2}{*}{14999513} & \multirow{2}{*}{3068195} & idealloc & 1638008 & 0.048 \\ \cline{7-9}
        & & & & & & real & 1622136 & 0.036 \\ \cline {2-9}
        & \multirow{2}{*}{jemalloc} & \multirow{2}{*}{1799744} & \multirow{2}{*}{786432} & \multirow{2}{*}{17623959} & \multirow{2}{*}{3565175} & idealloc & 1806344 & 0.039 \\ \cline{7-9}
        & & & & & & real & 8480992 & 0.084 \\ \cline {2-9}
        & \multirow{2}{*}{mimalloc} & \multirow{2}{*}{2455104} & \multirow{2}{*}{1048576} & \multirow{2}{*}{24551040} & \multirow{2}{*}{4855554} & idealloc & 2459656 & 0.024 \\ \cline{7-9}
        & & & & & & real & 4652672 & 0.032 \\ \cline {2-9}
        & \multirow{2}{*}{snmalloc} & \multirow{2}{*}{2520656} & \multirow{2}{*}{1048576} & \multirow{2}{*}{25206560} & \multirow{2}{*}{4975939} & idealloc & 2528784 & 0.024 \\ \cline{7-9}
        & & & & & & real & 16728144 & 0.054 \\ \hline
        \multirow{8}{0.1\textwidth}{java-scimark2 (11261)} & \multirow{2}{*}{glibc} & \multirow{2}{*}{5358936} & \multirow{2}{*}{528376} & \multirow{2}{*}{50939916} & \multirow{2}{*}{9608547} & idealloc & 8659120 & 0.174 \\ \cline{7-9}
        & & & & & & real & 6711904 & 0.167 \\ \cline {2-9}
        & \multirow{2}{*}{jemalloc} & \multirow{2}{*}{5428672} & \multirow{2}{*}{524288} & \multirow{2}{*}{51572384} & \multirow{2}{*}{9720874} & idealloc & 9108328 & 0.222 \\ \cline{7-9}
        & & & & & & real & 22833600 & 0.174 \\ \cline {2-9}
        & \multirow{2}{*}{mimalloc} & \multirow{2}{*}{5526976} & \multirow{2}{*}{524288} & \multirow{2}{*}{52506272} & \multirow{2}{*}{9885714} & idealloc & 7639624 & 0.149 \\ \cline{7-9}
        & & & & & & real & 11009664 & 0.180 \\ \cline {2-9}
        & \multirow{2}{*}{snmalloc} & \multirow{2}{*}{5953440} & \multirow{2}{*}{524288} & \multirow{2}{*}{56557680} & \multirow{2}{*}{10598910} & idealloc & 8272960 & 0.147 \\ \cline{7-9}
        & & & & & & real & 16728064 & 0.190 \\ \hline
    \end{tabular}
\end{table*}

\begin{table*}
    \caption{(continued from Table \ref{tab:frag1})}
    \label{tab:frag2}
    \centering
    \begin{tabular}{|p{0.1\textwidth}|l|c|c|c|c|l|c|c|}
        \hline
        \textbf{Workload (\#jobs)} & \textbf{Allocator} & $\mathbf{L}$ & $\mathbf{h_{max}}$ & \multicolumn{1}{p{0.03\textwidth}|}{\textbf{Robson bound}} & \multicolumn{1}{p{0.03\textwidth}|}{\textbf{BA bound}} & \textbf{Setting} & \textbf{Makespan} & \textbf{Fragmentation} \\ \hline
        \multirow{8}{0.1\textwidth}{compress-gzip (22963)} & \multirow{2}{*}{glibc} & \multirow{2}{*}{318304} & \multirow{2}{*}{32832} & \multirow{2}{*}{2387728} & \multirow{2}{*}{572349} & idealloc & 627600 & 0.465 \\ \cline{7-9}
        & & & & & & real & 339776 & 0.210 \\ \cline {2-9}
        & \multirow{2}{*}{jemalloc} & \multirow{2}{*}{383400} & \multirow{2}{*}{40960} & \multirow{2}{*}{2937213} & \multirow{2}{*}{690939} & idealloc & 693608 & 0.408 \\ \cline{7-9}
        & & & & & & real & 1265472 & 0.450 \\ \cline {2-9}
        & \multirow{2}{*}{mimalloc} & \multirow{2}{*}{383400} & \multirow{2}{*}{40960} & \multirow{2}{*}{2937213} & \multirow{2}{*}{690939} & idealloc & 774520 & 0.397 \\ \cline{7-9}
        & & & & & & real & 2567808 & 0.306 \\ \cline {2-9}
        & \multirow{2}{*}{snmalloc} & \multirow{2}{*}{383568} & \multirow{2}{*}{40960} & \multirow{2}{*}{2938500} & \multirow{2}{*}{691222} & idealloc & 816464 & 0.416 \\ \cline{7-9}
        & & & & & & real & 16744448 & 0.507 \\ \hline
        \multirow{8}{*}{dcraw (63)} & \multirow{2}{*}{glibc} & \multirow{2}{*}{81561656} & \multirow{2}{*}{81526776} & \multirow{2}{*}{1071751586} & \multirow{2}{*}{171607522} & idealloc & 81588600 & 0.000 \\ \cline{7-9}
        & & & & & & real & 81564744 & 0.000 \\ \cline {2-9}
        & \multirow{2}{*}{jemalloc} & \multirow{2}{*}{83922104} & \multirow{2}{*}{83886080} & \multirow{2}{*}{1104495793} & \multirow{2}{*}{176573935} & idealloc & 83931320 & 0.000 \\ \cline{7-9}
        & & & & & & real & 107294064 & 0.000 \\ \cline {2-9}
        & \multirow{2}{*}{mimalloc} & \multirow{2}{*}{83922104} & \multirow{2}{*}{83886080} & \multirow{2}{*}{1104495793} & \multirow{2}{*}{176573935} & idealloc & 251694312 & 0.333 \\ \cline{7-9}
        & & & & & & real & 285212288 & 0.000 \\ \cline {2-9}
        & \multirow{2}{*}{snmalloc} & \multirow{2}{*}{134253760} & \multirow{2}{*}{134217728} & \multirow{2}{*}{1812425760} & \multirow{2}{*}{282476244} & idealloc & 134262976 & 0.000 \\ \cline{7-9}
        & & & & & & real & 301798384 & 0.000 \\ \hline
        \multirow{8}{0.1\textwidth}{system-libxml2 (200512)} & \multirow{2}{*}{glibc} & \multirow{2}{*}{153984} & \multirow{2}{*}{72720} & \multirow{2}{*}{1243425} & \multirow{2}{*}{306717} & idealloc & 172784 & 0.267 \\ \cline{7-9}
        & & & & & & real & 200512 & 0.371 \\ \cline {2-9}
        & \multirow{2}{*}{jemalloc} & \multirow{2}{*}{169264} & \multirow{2}{*}{81920} & \multirow{2}{*}{1381357} & \multirow{2}{*}{337742} & idealloc & 195376 & 0.265 \\ \cline{7-9}
        & & & & & & real & 891584 & 0.551 \\ \cline {2-9}
        & \multirow{2}{*}{mimalloc} & \multirow{2}{*}{169264} & \multirow{2}{*}{81920} & \multirow{2}{*}{1381357} & \multirow{2}{*}{337742} & idealloc & 193648 & 0.268 \\ \cline{7-9}
        & & & & & & real & 3092480 & 0.420 \\ \cline {2-9}
        & \multirow{2}{*}{snmalloc} & \multirow{2}{*}{218464} & \multirow{2}{*}{131072} & \multirow{2}{*}{1856944} & \multirow{2}{*}{442691} & idealloc & 202521 & 0.199 \\ \cline{7-9}
        & & & & & & real & 945232 & 0.490 \\ \hline
        \multirow{8}{0.1\textwidth}{tjbench (34618)} & \multirow{2}{*}{glibc} & \multirow{2}{*}{7607792} & \multirow{2}{*}{7307256} & \multirow{2}{*}{86732245} & \multirow{2}{*}{15959249} & idealloc & 7637200 & 0.007 \\ \cline{7-9}
        & & & & & & real & 7612448 & 0.004 \\ \cline {2-9}
        & \multirow{2}{*}{jemalloc} & \multirow{2}{*}{7674960} & \multirow{2}{*}{7340032} & \multirow{2}{*}{87522768} & \multirow{2}{*}{16094960} & idealloc & 7693960 & 0.006 \\ \cline{7-9}
        & & & & & & real & 13691920 & 0.006 \\ \cline {2-9}
        & \multirow{2}{*}{mimalloc} & \multirow{2}{*}{7707728} & \multirow{2}{*}{7340032} & \multirow{2}{*}{87896444} & \multirow{2}{*}{16158532} & idealloc & 7726728 & 0.006 \\ \cline{7-9}
        & & & & & & real & 8830976 & 0.005 \\ \cline {2-9}
        & \multirow{2}{*}{snmalloc} & \multirow{2}{*}{8756320} & \multirow{2}{*}{8388608} & \multirow{2}{*}{100697680} & \multirow{2}{*}{18365011} & idealloc & 8775312 & 0.005 \\ \cline{7-9}
        & & & & & & real & 16711728 & 0.006 \\ \hline
    \end{tabular}
\end{table*}

\subsubsection{Producing new $\epsilon$ values}
A few simple heuristics oversee the process that recompute Corollary 17's error parameter:

\begin{itemize}
    \item in the first iteration, follow the formula in Section \ref{sec:c17}
    \item if the last value did not box \textit{any} jobs, increase $\epsilon$ by 10\%
    \item if the last value boxed \textit{more} jobs than the previous one, keep it as a bottom limit. There is no point in trying smaller values in the future
    \item if, after its increase, $\epsilon$ exceeds two times the bottom limit, start from the bottom again. There is no point in trying too big values either
\end{itemize}

\subsubsection{Help signals}
Apart from the error signal, we made use of two more that proved necessary. The first has already been mentioned: we keep track of how many jobs from the original input set managed to get boxed before failure. The second is discussed below.

\subsection{Bypassing Theorem 2 edge cases with critical time injection}
Just as Theorem 16 depends on whether $X_S$ contains any jobs in order to converge, Theorem 2 depends on $R$ being non-empty (see Figure \ref{fig:t2}).

The sorting operation of Theorem 2 uses a liveness criterion: jobs that are live during at least one critical time go in $R$. We noticed that, for some edge cases, this does not hold for any job in the input. In theory this should not happen: it is nowhere implied in the BA paper that one should worry whether $R$ is empty or not. But in practice we must. In our opinion, this is owed to the ``random moment'' that initializes the component's bounding intervals (see Section \ref{sec:t2}). Given that time is measured in allocated bytes, the range is vast enough to allow such edge cases to pop up.

Thus, upon detecting an empty $R$, apart from an error signal \textit{a different} moment is returned, at which the problematic input does contain live jobs. This is the only occasion where the error is not propagated all the way back to the entry point, because it makes sense to retry Theorem 2 with a bounding intervals vector injected with the moment returned. If the second attempt fails as well, we allow our implementation to go rogue and use an appropriate critical time whenever it stumbles on the edge case described \textit{without} returning.\footnote{Yes, this is as much cheating as it looks. We expect and celebrate future efforts that provide a final answer, stemming from more principled methods than trial-and-error.}

\subsection{Processing boxed jobs to retrieve the final placement}
Upon convergence BA produces a set of boxes of identical height. These contain boxes, which contain boxes, which contain other boxes, and so on until at the heart of each box, where a subset of the initial jobs resides. At this point everything has been boxed, but nothing has been placed. As Figure \ref{fig:safe} shows in its right half, two tasks remain:

\begin{itemize}
    \item derive an optimal outer box placement via IGC
    \item unbox placed boxes recursively until original jobs are found. At each unboxing stage, place children, i.e., contained boxes, according to their parent's placement
\end{itemize}

An important detail that goes unnoticed in the original publication is that the placement that emerges after unboxing the last job is \textit{very sparse}, probably owed to the recursive scaling imposed by Corollary 15. To tighten said placement, a very simple algorithm can be followed: (i) order jobs by increasing address, (ii) place the first job at address zero, and (iii) keep a vector of all traversed jobs. Traverse all remaining jobs, each time checking the traversed vector for temporally coinciding jobs with the current one. If such jobs exist, place the job right on top of the tallest coinciding one. Else place it at address zero. Keep traversing until done.

\subsection{A note on performance and scalability}
BA's recursive nature tends to drive even a server with 32 GiB DRAM to out-of-memory errors for workloads containing a mere 20 thousand jobs. This number is at least two orders of magnitude smaller than realistic workload sizes. Note, however, that job count is not the definitive factor to predict OOM killers: as will be shown later (Tables \ref{tab:frag1} and \ref{tab:frag2}), we managed to pull complete BA runs on 80K and 200K job workloads. This leads us to conclude that more fine-grain characteristics, such as the size distributions of jobs as well as their placement in time (controlled by program behavior), are more relevant. We have not come up with a criterion to differentiate between tractable and intractable cases.

\section{Results and discussion}
\label{sec:res}
We run experiments on a \texttt{x86\_64} commodity server running Ubuntu 20.04 LTS. We collected placement data from a pool of applications linked to a pool of state-of-the-art allocators, and then fed that data to BA. All benchmarks come from the Single-Threaded Tests collection on openbenchmarking.org.\footnote{\href{https://openbenchmarking.org/suite/pts/single-threaded}{https://openbenchmarking.org/suite/pts/single-threaded}} We used the Phoronix Test Suite\footnote{\href{https://www.phoronix-test-suite.com/}{https://www.phoronix-test-suite.com/}} to install and run the applications. A complete record of our measurements can be found at Tables \ref{tab:frag1} and \ref{tab:frag2}.

All results shown are products of trace-based simulation. Linux kernel-side virtual-to-physical mapping decisions do not matter in this context: the underlying assumption is that \textit{all mappings are contiguous}. All evaluation regards placement on the virtual address space. The degree to which such decisions affect main memory is established in this paper's sister work (see Section \ref{sec:fatality} for details).

\subsection{Bounds comparison}
We have two reasons to compare placements with theoretical bounds: on the allocators' side, it is a good chance to see how they perform with respect to Robson's worst-case lower bound, defined as $\frac{1}{2}L\cdot log_2{h_{max}}$ for an ``optimal strategy''~\cite{robson74,robson1977worst}.  On BA's side, abidance to the authors' theoretical upper bound, that is, $M \leq [1 + {O((h_{max}/L)}^{1/7})]L$, is a reliable indicator of our implementation's functional correctness--which is more than desired, given the multitude of ad-hoc fixes that we applied toward convergence. Both comparisons are depicted at Figure \ref{fig:bounds}. 

We see that allocators outperform Robson's worst-case bound in 30 out of 32 cases--despite employing sparse addressing. Note that we compute the makespan via subtracting the smallest address used from the largest one, and that we always operate on virtual addresses. Consequently, even regions between the ``endpoint'' blocks that may never have gotten mapped to physical memory are accounted for--and thus sparse addressing \textit{does} affect the reported makespan. This result comes in line with earlier work producing similar conclusions in the RSS domain~\cite{fragsolved, reconsider}. 

With respect to BA, we get a much tighter upper bound. $75\%$ of the time, our implementation does indeed conform to $M \leq [1 + 2\cdot ({h_{max}/L)}^{1/7})]L$.

Of particular note is \texttt{glibc}'s terrific performance against the Robson lower bound across all workloads. Upon inspection of visualized placement files like the one shown in Figure \ref{fig:temple}, we found out that \texttt{glibc} almost never uses sparse addressing within each spawned memory mapping--as opposed to more ``modern'' implementations such as \texttt{mimalloc}. We attribute its very small makespan to this observation.

\subsection{Allocators vs. BA}
Figure \ref{fig:versus} displays the empirical evidence regarding allocators' performance against BA. Quite a few interesting remarks can be made. This being an intellectual abstract, we shall not try and fully explain the data shown. We invest our best of efforts, however, to provide a first interpretation.

\begin{enumerate}
    \item $75\%$ of the time, BA outperforms allocators as regards makespan. This is expected, given BA's offline, make-\\span-optimizing nature.
    
    A less obvious detail is that, trend-wise, Figure \ref{fig:versus}'s left half largely follows Figure \ref{fig:bounds}'s left half. In other words, allocators' makespan versus BA is proportionate to allocators' makespan versus Robson's lower bound. This could hint at BA makespan being more appropriate as a lower bound than Robson's values. But we leave said question open for this paper
    
    \item \texttt{glibc} yields makespan that is equivalent to or lower than BA in 7 out of 8 workloads. We have already mentioned \texttt{glibc}'s observed reluctance for sparse placements, but still the fact that it is such a fierce competitor to BA is surprising. 
    
    Again, Figure \ref{fig:bounds} carries some information about this phenomenon: BA is defeated to the greatest degree in workloads where it fared the worst regarding its own upper bound (\texttt{bullet}, \texttt{java-scimark2}, \texttt{compress-\\gzip})
    
    \item The most surprising finding is that BA outperforms or is equivalent to allocators with respect to page-local fragmentation more than $65\%$ of the time. 
    
    How good BA behaves seems once again linked to how its makespan compares to its upper bound (Figure \ref{fig:bounds}, right half). \texttt{bullet} and \texttt{dcraw-mimalloc} are the most obvious cases, but closer inspection illuminates this observation's uniformity across all workloads
\end{enumerate}

We hope the above points to be sufficient as inspiration for future work. If 2DBP fragmentation does indeed affect RSS, then one can imagine profile-guided optimization of individual workloads based on ``optimal'' placements computed by BA. Analogous efforts targeting makespan could be attempted in domains where it makes sense.

\section{Related Work}
\label{sec:rw}
Wilson et al. have written the seminal treatment on DSA and the central role of fragmentation~\cite{wilson1995surv}. Johnstone and Wilson conduct the first study of RSS-based fragmentation definitions~\cite{fragsolved}. Berger et al. show that modern allocators perform acceptably well with respect to RSS-based fragmentation~\cite{reconsider}. Maas et al. propose a novel fragmentation definition incorporating chances of immediate memory reuse~\cite{realfrag}. Powers et al. and Maas et al. contribute notably unorthodox ways to deal with fragmentation~\cite{mesh, learnalloc}.

On the theoretical side, Robson has computed general worst-case fragmentation bounds for any policy~\cite{robson71, robson74}, as well as tighter bounds for the best fit and first fit policies~\cite{robson1977worst}. Optimal placement is reported as NP-hard by Garey and Johnson~\cite{garey_jognson_1979}. Chrobak and {\'S}lusarek formulate DSA as a 2DBP instance~\cite{chrobak1988some}. Given our focus on 2DBP, we do not mention other formulations such as graph coloring~\cite{KIERSTEAD1991231}.

\section{Conclusion}
\label{sec:end}
This paper brings into focus the theoretical branch of literature dealing with dynamic memory allocation, envisioning to exploit its state-of-the-art to the advantage of real-world systems. It is built on top of work which proves that two-dimensional rectangle bin packing is an informative representation of workload-allocator interaction. We extend that work by (i) implementing the best known bin packing algorithm suitable for modeling dynamic memory allocation and (ii) comparing its products, both in terms of makespan and fragmentation, with four modern allocators. Our demonstration aspires to spark further interest towards crossbreeds of theoretical and practical memory memory management.

\section*{Acknowledgements}
This work would not exist were it not for the original BA paper published by Buchsbaum et al. in 2003.~\cite{buchsbaum} We thus thank Adam Buchsbaum, Howard Karloff, Claire Mathieu, Nick Reingold and Mikkel Thorup for their contribution. Moreover we thank Paul Wilson, Mark Johnstone, Michael Neely and David Boles for inspiring us to study DSA from first principles.~\cite{wilson1995surv}

We owe all improvements on our initially submitted version to the feedback received from ISMM's Reviewers and Shepherd. We are particularly thankful to the Shepherd for guiding us through the last mile, as well as to Professor Erez Petrank for instantly resolving any inquiry. We thank the ISMM Organizing and Program Committees for allowing us an ideal space to showcase our work.

This research was supported by the Hellenic Foundation for Research and Innovation (HFRI) under the 3rd Call for HFRI PhD Fellowships (Fellowship Number: 61/512200), and by the European Union's Horizon 2020 research and innovation programme under grant agreement No. 101021274.
\balance
\bibliographystyle{ACM-Reference-Format}
\balance
\bibliography{sample-base}
\balance
\end{document}